\documentclass{epl}
\usepackage{amssymb,graphicx,psfrag,amsmath}

\title{Magnetotransport in d-wave density waves}
\shorttitle{Magnetotransport in d-wave density waves}
\author{Bal\'azs D\'ora\inst{1} \and Kazumi Maki\inst{2,3} 
\and Attila Virosztek\inst{1,4}}
\institute{\inst{1} Department of Physics, Budapest University of Technology and Economics,
 H-1521 Budapest, Hungary \\
\inst{2} Department of Physics and Astronomy, University of Southern California, Los Angeles
CA 90089-0484, USA \\
\inst{3} Max-Planck-Institut f\"ur Physik komplexer Systeme, N\"othnitzer Str. 38, D-01187 Dresden, 
Germany\\
\inst{4} Research Institute for Solid State Physics and Optics, P.O.Box 49,
H-1525 Budapest, Hungary}


\date{}

\begin{document}
\maketitle
\begin{abstract}
Angle dependent magnetoresistance (ADMR) and giant Nernst effect are hallmarks of unconventional density waves (UDW).
Here these transport properties for d-wave density wave (d-DW) are computed for quasi-two-dimensional systems. The present theory
describes ADMR observed in the pseudogap phase of Y$_{0.68}$Pr$_{0.32}$Ba$_2$Cu$_3$O$_7$ and CeCoIn$_5$ single crystals very satisfactorily.

\end{abstract}

\section{Introduction}

As is well known, many electronic systems like high $T_c$ cuprates, heavy fermion systems and organic 
conductors exhibit the pseudogap phenomenon or so-called non-Fermi liquid behaviour\cite{thalmeier}. 
Further, several people 
proposed that the pseudogap phase in the underdoped region of high $T_c$ cuprates is d-wave density 
wave\cite{capelluti,benfatto,nayak}. 
We have shown recently that the giant Nernst effect seen in the pseudogap region of LSCO, YBCO and 
Bi2212\cite{xu,wang1,wang2,capan} 
can be interpreted in terms of d-wave density wave (d-DW)\cite{capnernst}. Also the Pauli 
limiting 
behaviour of d-DW in the
c-axis resistance\cite{shibauchi,krusin} indicates that it is d-wave charge density wave and not d-wave 
spin 
density wave\cite{tesla}.

Also there are many parallels between organic superconductor $\kappa$-(ET)$_2$ salts\cite{pinteric}, heavy 
fermion 
superconductor CeCoIn$_5$ and high $T_c$ cuprate superconductors. In particular the quasi-two-dimensional 
Fermi surface, the layered structure, d-wave superconductivity\cite{izawa,izawa2} and the presence of 
pseudogap are noteworthy.
Very recently d-DW in the pseudogap phase of CeCoIn$_5$ was established through the giant Nernst 
effect\cite{bel,cecoin} and 
the angle dependent magnetoresistance\cite{tao}.

As noted by Nersesyan et al.\cite{Ners1,Ners2}, the quasiparticle spectrum in unconventional density wave 
is quantized in a 
magnetic field. This Landau quantization gives rise to the spectacular angle dependent 
magnetoresistance 
(ADMR) and giant Nernst effect as reviewed in Ref. \cite{mplb}. These are exploited to identify the 
presence of UDW in 
$\alpha$-(ET)$_2$ KHg(SCN)$_4$\cite{admrprl}, and Bechgaard salts (TMTSF)$_2$X with X=PF$_6$ and 
ReO$_4$\cite{kang,tmtsf}.

In spite of these successes earlier works are limited to quasi-one-dimensional systems. In this 
paper we shall first consider the Landau quantization of d-wave density waves.
Then the ADMR data on Y$_{0.68}$Pr$_{0.32}$Ba$_2$Cu$_3$O$_7$\cite{sandu} and in CeCoIn$_5$ are reanalyzed. 
Y$_{0.68}$Pr$_{0.32}$Ba$_2$Cu$_3$O$_7$ becomes superconducting at 55~K, above it pseudogap behaviour is 
seen.  
Also the gap 
structure of d-DW on the quasi-two-dimensional Fermi surface can be fully exploited in CeCoIn$_5$.

\section{Landau quantization in d-DW}

The quasiparticle energies in a d-wave density wave are determined from the poles of the Nambu Green's 
function\cite{nambu}:
\begin{equation}
G^{-1}(\omega,{\bf k})=\omega-\xi({\bf k})\rho_3-\eta({\bf k})-\Delta({\bf k})\rho_1,
\label{elso}
\end{equation}
where $\rho_i$'s are the Pauli's matrices operating on the spinor space. 
These energies are assumed to describe the system in the pseudogap regime, above 55~K in 
Y$_{0.68}$Pr$_{0.32}$Ba$_2$Cu$_3$O$_7$, although similar Green's function can be obtained in the 
superconducting region. The basic difference between the two stems from the fact that the spinor space in 
superconductors consists of $(c_{\bf k,\sigma},c^+_{\bf -k,-\sigma})$ while in DW $(c_{\bf 
k,\sigma},c_{\bf k-Q,\sigma})$.
For a d-wave density wave, we can 
further assume $\Delta({\bf k})=\Delta\cos(2\phi)$ or $\Delta\sin(2\phi)$ with $\tan(\phi)=k_y/k_x$,
\begin{gather}
\varepsilon({\bf k})=-2t(\cos(ak_x)+\cos(ak_y))-\mu,\\
\eta({\bf k})=\frac{\varepsilon({\bf k})+\varepsilon({\bf k-Q})}{2}=-\mu,\\
\xi({\bf k})=\frac{\varepsilon({\bf k})-\varepsilon({\bf k-Q})}{2},
\end{gather}
$\mu$ the chemical potential, which acts as imperfect nesting, $\bf Q$ is the best nesting vector. 
Since we are only interested in the low energy excitation around the nodes of gap, we can use a linearized 
version of the spectrum valid around these points as $\xi({\bf 
k})=v(k-k_\parallel)+\frac{v'}{c}\cos(ck_z)$ with $v$ and $v'$ the in-plane and c-axis Fermi velocities. 
Here dispersion perpendicular to the plane was also taken into account.
In the presence of magnetic field $\bf B$ within the x'-z plane, where $x'=\hat x\cos{\phi}+\hat y 
\sin(\phi)$ and tilted by $\theta$ from the z axis, the effect of magnetic field is introduced in eq. 
(\ref{elso}) through the Peierls-Onsager substitution ${\bf k}={\bf k}+e{\bf A}$ and
\begin{equation}
{\bf A}=B(y\cos(\phi)-x\sin(\phi)(\hat z\sin(\theta)-(\hat x\cos(\phi)+\hat y\sin(\phi))\cos(\theta)).
\end{equation}
Then the quasiparticle spectrum is determined by
\begin{gather}
E\Psi=\left\{\left[-veB\cos(\theta)(\cos(\phi)y-\sin(\phi)x)+\frac{v'}{c}\cos\left(ceB\sin(\theta)
(\cos(\phi)y-\sin(\phi)x)+\chi\right)\right]\rho_3-\right.\nonumber\\
-\left.\mu-iv_2\rho_1\partial_y\right\}\Psi,
\end{gather}
where $v_2/v=\Delta/E_F$ and we assume here $\Delta({\bf k})=\Delta\sin(2\phi)$ or d$_{xy}$-wave DW for 
simplicity. Further for CeCoIn$_5$ we have $c^2eH\sim 10^{-2}$ for H=1~T. Then $\cos(ceB\sin(\theta)
(\cos(\phi)y-\sin(\phi)x)+\chi)\simeq\pm ceB\sin(\theta)(\cos(\phi)y-\sin(\phi)x)$ for $\chi=\pm\pi/2$.
Then the solution is easily obtained following Weisskopf\cite{weisskopf,jackiw}. These are four branches of 
the 
quasiparticles; 
two branches around the Dirac cone [1,0,0] and others at [0,1,0] with
\begin{gather}
E^\pm_{1n}=\pm\sqrt{2neBv_2|v\cos(\theta)\cos(\phi)-v'\sin(\theta)||\cos(\phi)|}-\mu\label{egy}\\
E^\pm_{2n}=\pm\sqrt{2neBv_2|v\cos(\theta)\cos(\phi)+v'\sin(\theta)||\cos(\phi)|}-\mu\label{ketto}\\
E^\pm_{3n}=\pm\sqrt{2neBv_2|v\cos(\theta)\sin(\phi)-v'\sin(\theta)||\sin(\phi)|}-\mu\label{harom}\\
E^\pm_{4n}=\pm\sqrt{2neBv_2|v\cos(\theta)\sin(\phi)+v'\sin(\theta)||\sin(\phi)|}-\mu\label{negy}.
\end{gather}
Here $n=0$, $1$, $2$\dots. Except for the $n=0$ Landau level, each Landau level is double degenerated.
Also the corresponding Landau wavefunctions are readily constructed as in Ref. \cite{mplb}. In particular
\begin{equation}
\Psi_0\sim\exp\left[-\frac 12 eB\frac{v}{v_2}\left|\cos(\theta)\pm g 
\frac{\sin(\theta)}{\cos(\phi)}\right|(y\cos(\phi)-x\sin(\phi))^2\right]
\end{equation}
from the Dirac cone at [1,0,0] and 
\begin{equation}
\Psi_0\sim\exp\left[-\frac 12 eB\frac{v}{v_2}\left|\cos(\theta)\pm g
\frac{\sin(\theta)}{\sin(\phi)}\right|(y\cos(\phi)-x\sin(\phi))^2\right]
\end{equation}
from the one at [0,1,0].

From these Landau levels, the thermodynamics as well as the magnetotransport properties are readily 
deduced.

\section{Angular dependent magnetoresistance}

\begin{figure}[h!]
\psfrag{x}[t][b][1][0]{$\theta^\circ$}
\psfrag{y}[][t][1][0]{$\Delta\rho_{ab}/\rho_{ab}$}
\psfrag{t}[][][1][0]{T=52~K}
\psfrag{x2}[t][b][1][0]{$\theta^\circ$}
\psfrag{y2}[][t][1][0]{$\Delta\rho_{ab}/\rho_{ab}$}
\psfrag{t21}[][][1][0]{T=65~K}
\psfrag{t22}[][][1][0]{T=60~K}
\twofigures[width=7cm,height=7cm]{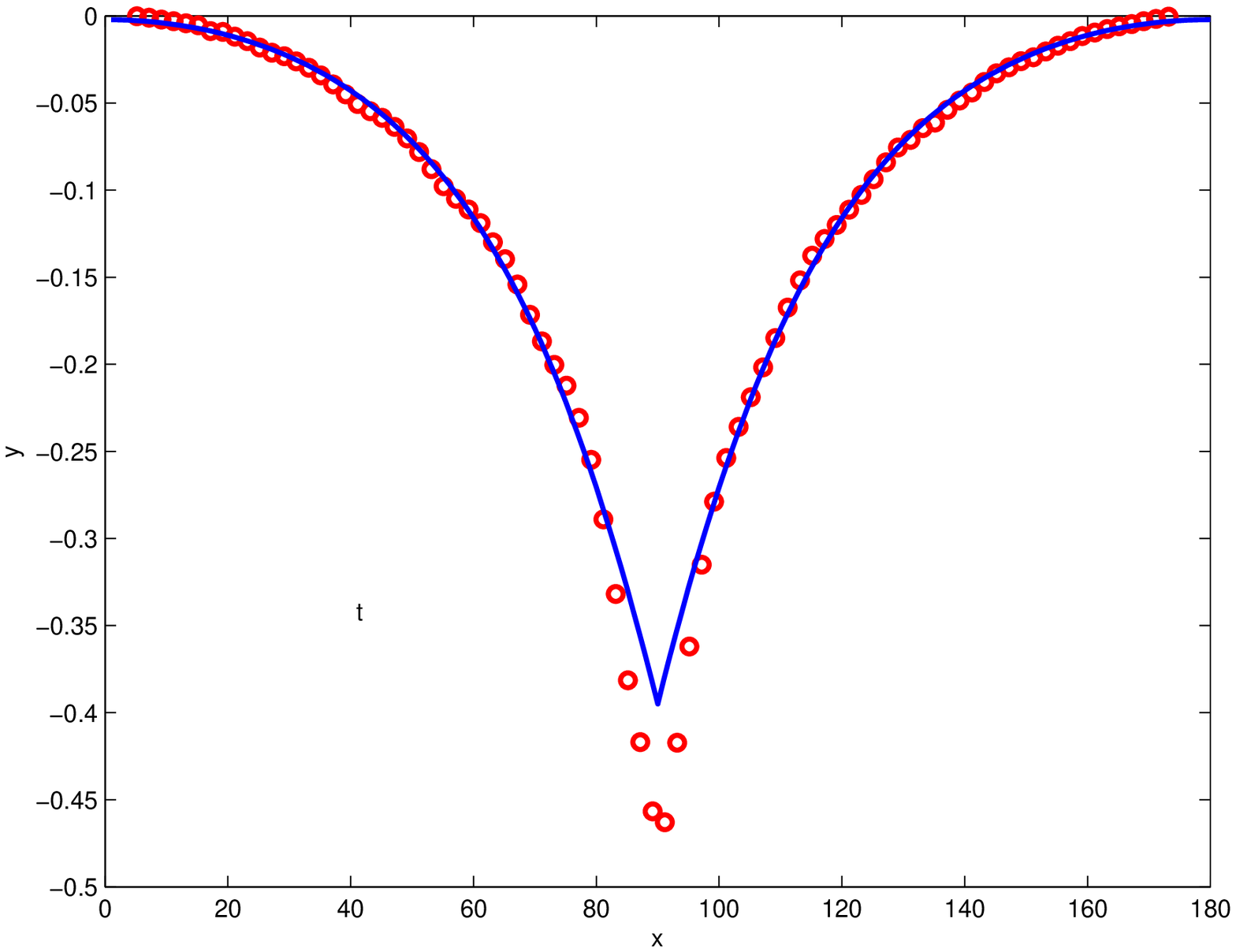}{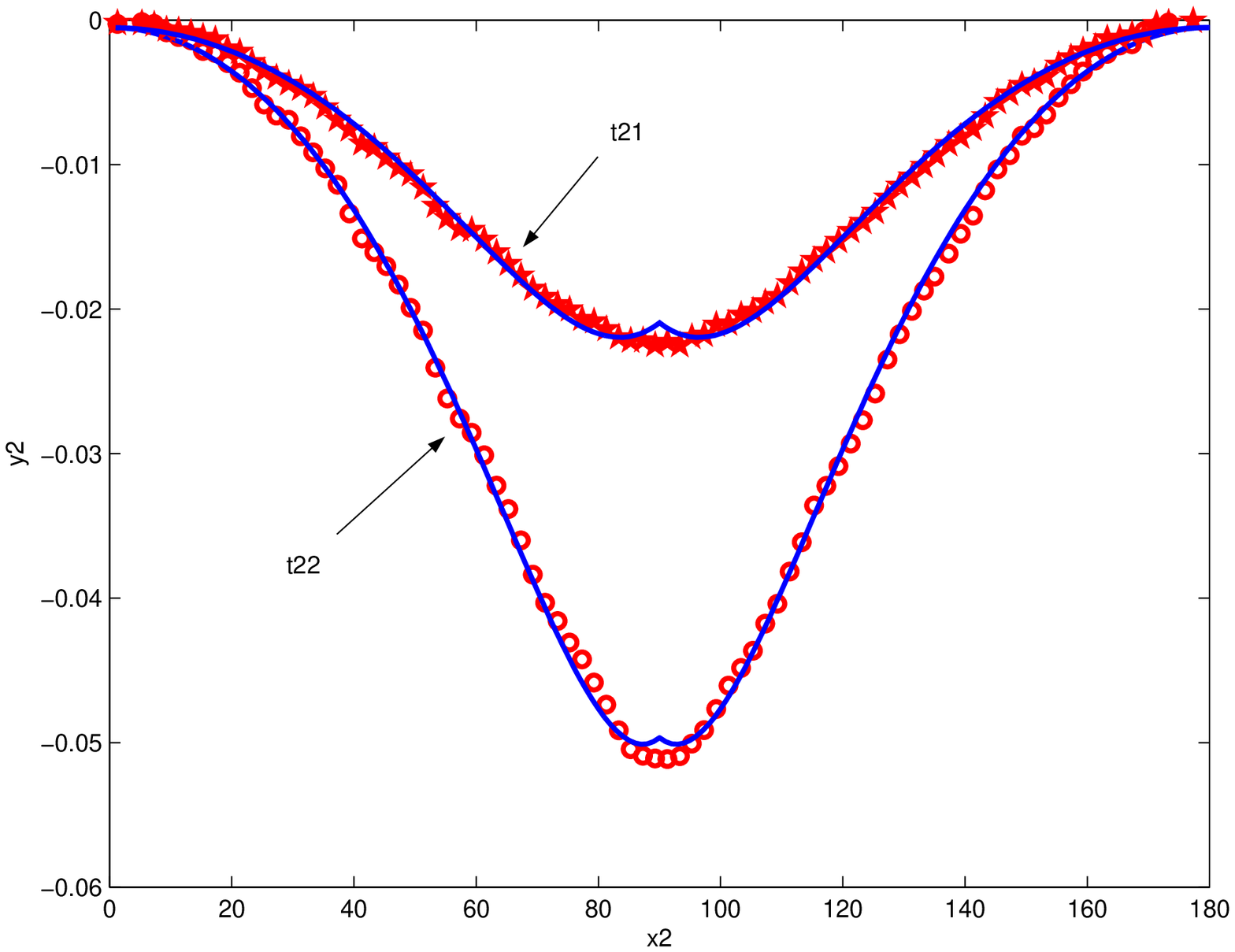}
\caption{The relative change of the in-plane magnetoresistance of 
Y$_{0.68}$Pr$_{0.32}$Ba$_2$Cu$_3$O$_7$\cite{sandu} is plotted as a function of angle $\theta$ at H=14~T 
for T=52~K. The solid line is fit based on eq. \ref{fittelo}.}
\label{rhoab1}
\caption{The relative change of the in-plane magnetoresistance of 
Y$_{0.68}$Pr$_{0.32}$Ba$_2$Cu$_3$O$_7$\cite{sandu}
is plotted as a function of angle 
$\theta$ at H=14~T 
for 60~K (circles), 65~K (pentagrams) together with our fit based on eq. \ref{fittelo}.}
\label{rhoab2}
\end{figure}

\begin{figure}[h!]
\psfrag{x3}[t][b][1][0]{$\theta^\circ$}
\psfrag{y3}[][t][1][0]{$\Delta\rho_{ab}/\rho_{ab}$}
\psfrag{t3}[][][1][0]{T=75~K}
\psfrag{x4}[t][b][1][0]{$\theta^\circ$}
\psfrag{y4}[][t][1][0]{$\Delta\rho_{ab}/\rho_{ab}$, T=105~K}
\psfrag{t4}[][][1][0]{T=105~K}
\twofigures[width=7cm,height=7cm]{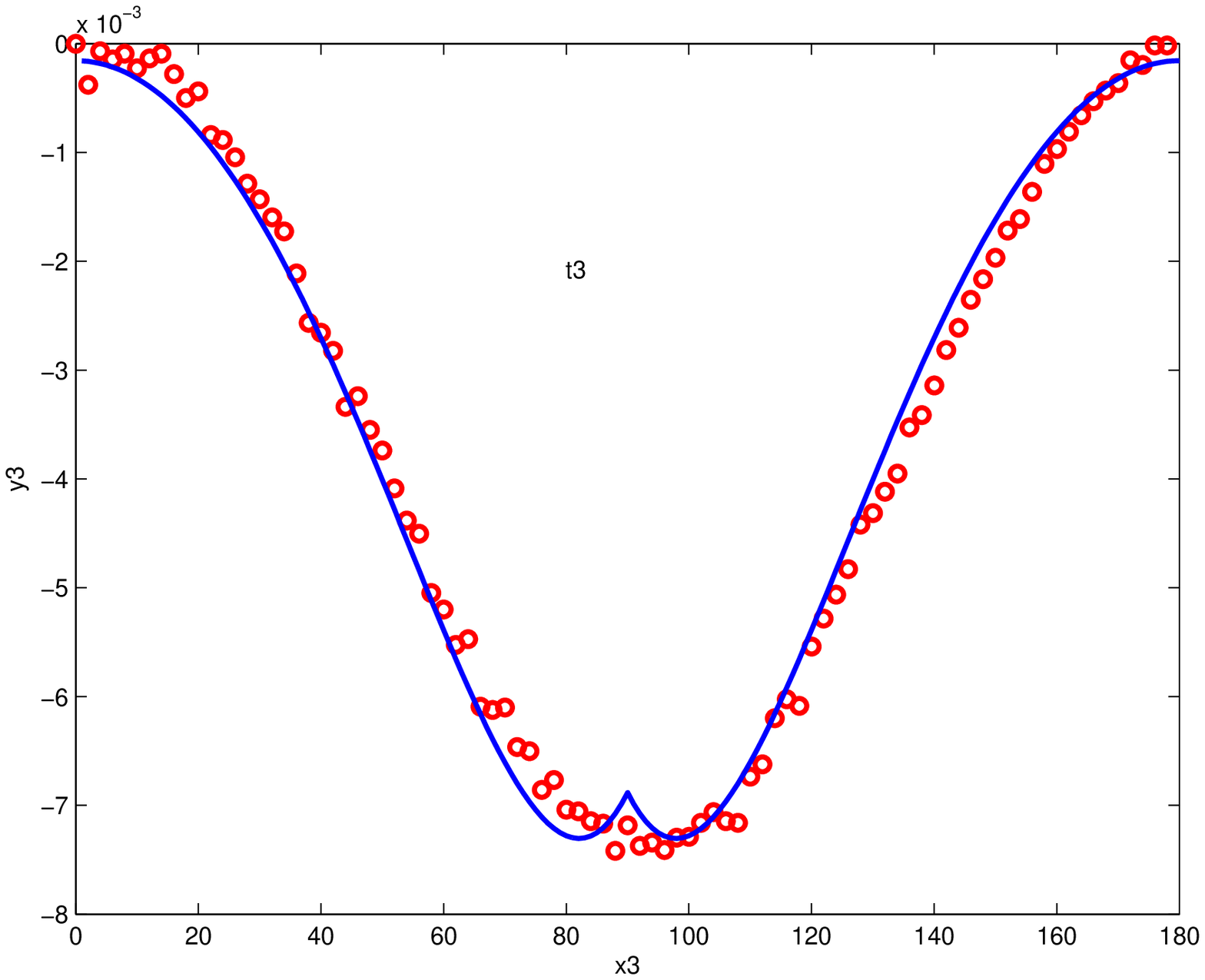}{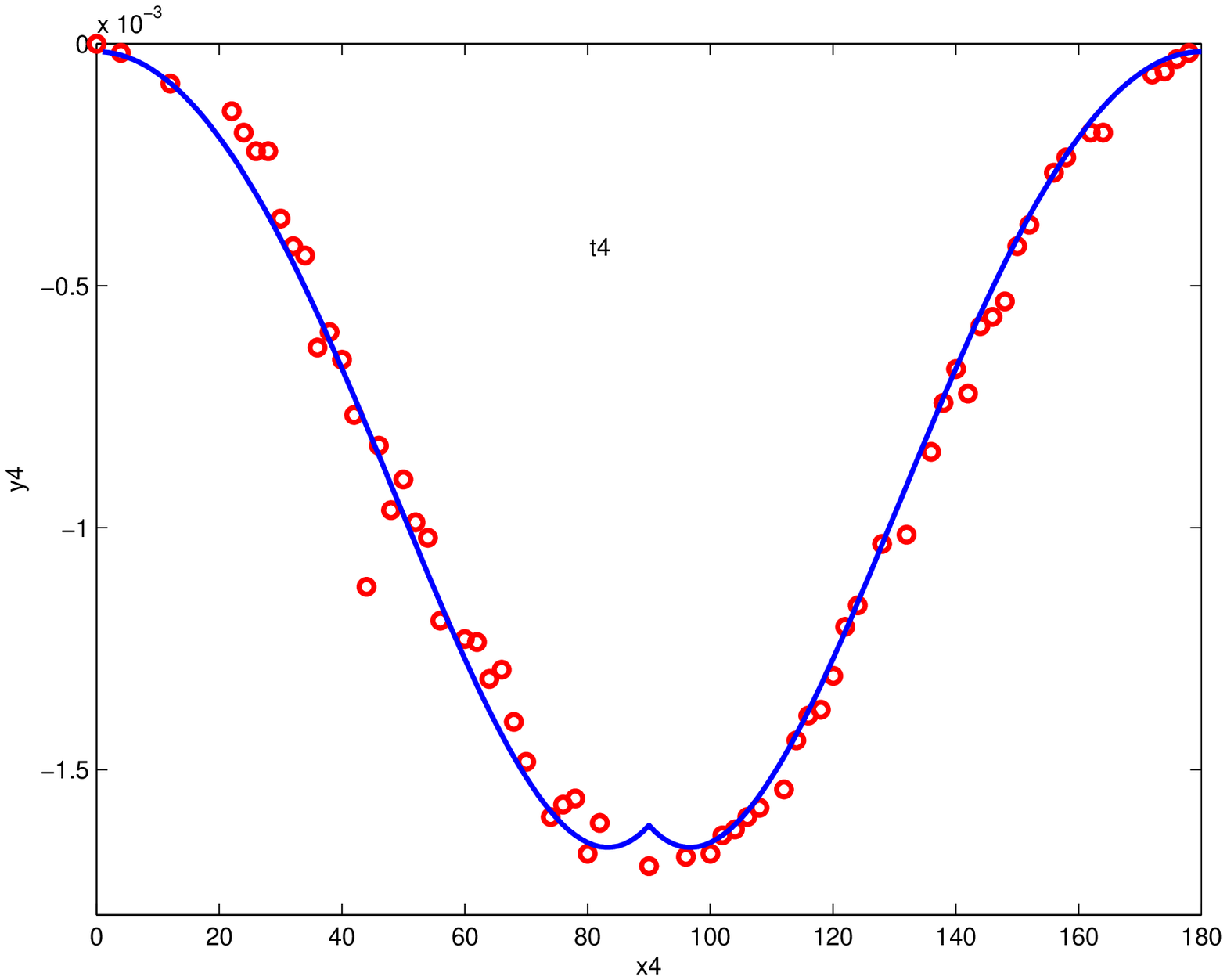}
\caption{The relative change of the in-plane magnetoresistance of 
Y$_{0.68}$Pr$_{0.32}$Ba$_2$Cu$_3$O$_7$\cite{sandu}
is plotted as a function of angle 
$\theta$ at H=14~T 
for T=75~K. The solid line is fit based on eq. \ref{fittelo}.}
\label{rhoab3}
\caption{The relative change of the in-plane magnetoresistance of 
Y$_{0.68}$Pr$_{0.32}$Ba$_2$Cu$_3$O$_7$\cite{sandu}
is plotted as a function of angle 
$\theta$ at H=14~T 
for 105~K together with our fit based on eq. \ref{fittelo}.}
\label{rhoab4}
\end{figure}

\begin{figure}[h!]
\psfrag{x}[t][b][1][0]{$\theta^\circ$}
\psfrag{y}[][t][1][0]{$\Delta\rho_{c}/\rho_{c}$, T=60~K, T=65~K}
\onefigure[width=7cm,height=7cm]{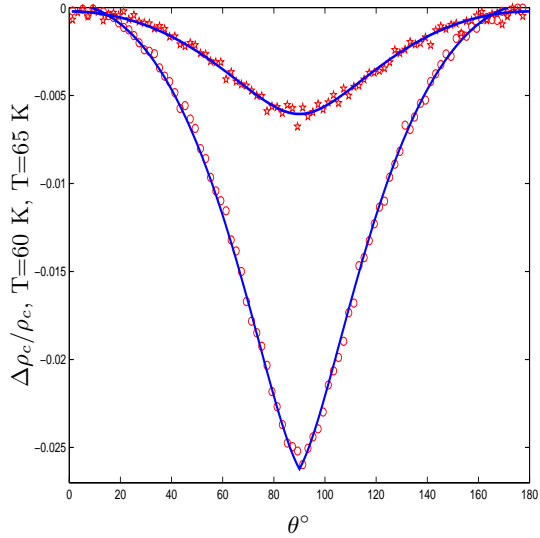}
\caption{The relative change of the c-axis magnetoresistance of Y$_{0.68}$Pr$_{0.32}$Ba$_2$Cu$_3$O$_7$\cite{sandu}
 is shown as a function 
of 
angle $\theta$ at H=14~T 
for 60~K (circles) and 65~K (pentagrams). The solid line represents our fit based on eq. \ref{fittelo}.}
\label{rhoc}
\end{figure}

We shall limit ourselves to  two limiting cases:\\
A. $\sigma_{xx}$ in a magnetic field in the z-y plane.\\
Here we take the angle $\bf B$ makes from the c-axis is $\theta$. Then the quasiparticle spectra are given 
by eqs. \ref{egy}-\ref{negy} with $\phi=\pi/2$.
So $E^\pm_{1n}$ and  $E^\pm_{2n}$ reduces to $-\mu$, the same as the $n=0$ Landau levels, while  
$E^\pm_{13n}$ and  $E^\pm_{4n}$ are given by
\begin{gather}
E^\pm_{3n}=\pm\sqrt{2neBv_2|v\cos(\theta)-v'\sin(\theta)|}-\mu\label{coll1}\\
E^\pm_{4n}=\pm\sqrt{2neBv_2|v\cos(\theta)+v'\sin(\theta)|}-\mu\label{coll2}.
\end{gather}
Then the electric conductivity is given by 
\begin{equation}
\sigma(B,\theta)=\sum\limits_n\sigma_n\textmd{sech}^2\left(\frac 12 \beta E_n\right),
\label{cond}
\end{equation}
where $E_n$'s are the energy of all the Landau levels. The above equation can be obtained by 
following the reasoning of ref. \cite{abrikosov}.
This expression is somewhat different from the one we proposed earlier\cite{mplb}, but we think more 
appropriate 
when $\mu\neq 0$. When $\beta|E_{31}|\gg1$ and  $\beta|E_{41}|\gg1$, only the lowest levels contribute 
significantly, and eq. \ref{cond} is simplified as
\begin{gather}
\sigma(B,\theta)=6\sigma_0\textmd{sech}^2\left(\frac 
12\zeta_0\right)+2\sigma_1\left(\textmd{sech}^2\left(\frac 
12(x_1-\zeta_0)\right)+\textmd{sech}^2\left(\frac 12(x_2-\zeta_0)\right)\right.+\nonumber \\
\left.\textmd{sech}^2\left(\frac 12(x_1+\zeta_0)\right)+
\textmd{sech}^2\left(\frac 
12(x_2+\zeta_0)\right)\right)=48\sigma_0\left(1+\cosh(\zeta_0)\right)^{-1}+\nonumber\\
+8\sigma_1\left[\frac{1+\cosh(x_1)\cosh(\zeta_0)}{(\cosh(x_1)+\cosh(\zeta_0))^2}+
\frac{1+\cosh(x_2)\cosh(\zeta_0)}{(\cosh(x_2)+\cosh(\zeta_0))^2}
\right],
\label{fittelo}
\end{gather}
where $\zeta_0=\beta\mu$, $x_1=\beta\sqrt{2eBv_2|v\cos(\theta)-v'\sin(\theta)|}$ and 
$x_2=\beta\sqrt{2eBv_2|v\cos(\theta)+v'\sin(\theta)|}$. In figs. \ref{rhoab1}-\ref{rhoc}, we show the 
in-plane and out of plane angle dependent 
magnetoresistance data from Y$_{0.68}$Pr$_{0.32}$Ba$_2$Cu$_3$O$_7$ by Sandu et al.\cite{sandu} together 
with our 
fitting based on eq. \ref{fittelo}. At 52~K, the applied magnetic field (14~T) can be strong enough to 
destroy superconductivity and drive the systems into the pseudogap regime, while at 105~K the presence of 
pseudogap is still felt.

\begin{figure}[h!]
\psfrag{x}[t][b][1][0]{$\phi^\circ$}
\psfrag{y}[][t][1][0]{$\sigma_{zz}$}
\onefigure[width=7cm,height=7cm]{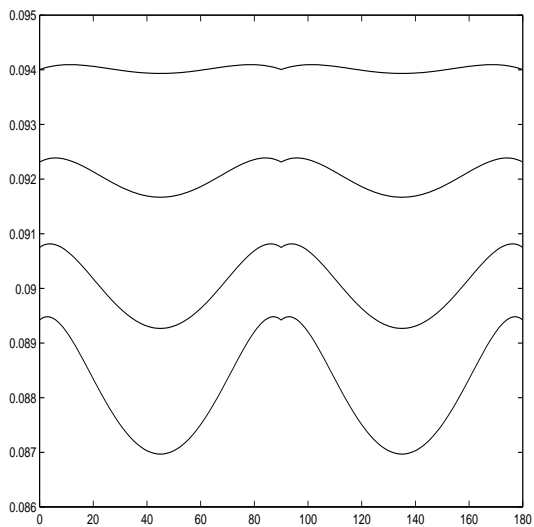}
\caption{The predicted $\phi$ dependence of $\sigma_{zz}$ in CeCoIn$_5$ is plotted for H=4~T, 6~T, 8~T and 10~T (from 
bottom to top).}
\label{sigmaphi}
\end{figure}

In high $T_c$ cuprates, $v'/v\ll 1$, so we have assumed $x_1=x_2$. From the fittings, using the universal 
Fermi velocity\cite{fermivel} $v=2.3\times 
10^7$~cm/s, we obtain $v_2=1.6\times 10^6$~cm/s and $\mu=40-60$~K. Since $E_F\sim 5000$~K in high $T_c$ 
cuprate we extract $\Delta=360$~K for 
Y$_{0.68}$Pr$_{0.32}$Ba$_2$Cu$_3$O$_7$ from $v_2/v=\Delta/E_F$. This value is consistent with a recent 
c-axis optical conductivity data on underdoped YBCO\cite{pimenov}.
Here $\Delta$ is 
the maximal gap of d-DW in 
the non-Fermi liquid or pseudogap phase of Y$_{0.68}$Pr$_{0.32}$Ba$_2$Cu$_3$O$_7$. 
The small deviations of the theory from the experimental results around $\theta=90^\circ$ originates from
the collapse of all the Landau levels in the $v'/v\ll 1$ limit, as seen from eqs. \ref{coll1} and 
\ref{coll2}.\\
B. ADMR in a magnetic field in the x-y plane.\\
For a d-DW it is of crucial importance to see if d$_{xy}$-wave density wave or d$_{x^2-y^2}$-wave 
density waves are realized.  For the d-DW in the pseudogap phase in high $T_c$ cuprates the angle resolved 
photoemission spectra (ARPES)\cite{ding} tells it is d$_{x^2-y^2}$-wave in high $T_c$ cuprates. On 
the 
other hand 
d$_{xy}$-wave appears to be more consistent with CeCoIn$_5$, as suggested by Aoki et al.\cite{aoki}. 
$\sigma_{zz}(B,\phi)$ in a rotating field within the a-b plane is given by
\begin{gather}
\sigma_{zz}(B,\phi)=16\sigma_0\left(1+\cosh(\zeta_0)\right)^{-1}
+8\sigma_1\left[\frac{1+\cosh(x_1)\cosh(\zeta_0)}{(\cosh(x_1)+\cosh(\zeta_0))^2}+
\frac{1+\cosh(x_2)\cosh(\zeta_0)}{(\cosh(x_2)+\cosh(\zeta_0))^2}
\right],
\label{condphi}
\end{gather}
where $x_1=\beta\sqrt{2eBv_2v'|\sin(\phi)|}$ and  $x_1=\beta\sqrt{2eBv_2v'|\cos(\phi)|}$
In fig. \ref{sigmaphi} the $\phi$ dependence of eq. \ref{condphi} is shown for 
parameters 
typical for CeCoIn$_5$\cite{tao}. Here we assumed 
d$_{xy}$-wave DW. For d$_{x^2-y^2}$-wave DW, the same expression applies if we shift $\phi$ to 
$\phi+\pi/4$. By varying the parameters, the small dip at 90$^\circ$ can be sharpened,  and 
the broad bump at 45$^\circ$ can be weakened, but these two features always remain present.

\section{Conclusion}

We have extended the early analysis of the Landau quantization\cite{Ners1,Ners2} to d-wave density wave. 
Then the 
quasiparticle spectrum in a magnetic field describes the ADMR in 
Y$_{0.68}$Pr$_{0.32}$Ba$_2$Cu$_3$O$_7$\cite{sandu} and 
in CeCoIn$_5$\cite{tao} very well. Though the present analysis suggests d$_{xy}$-wave density wave in 
CeCoIn$_5$, it 
is possible to discriminate d$_{x^2-y^2}$-wave and d$_{xy}$-wave DW through the ADMR where the magnetic 
field is rotated within the a-b plane.
Also the present results will be used for further exploration of d-wave DW in CeCu$_2$Si$_2$, 
URu$_2$Si$_2$, $\kappa$-(ET)$_2$ salts and many other compounds.

We are thankful to V. Sandu and C. C. Almasan for helpful discussion and for providing us with the data of 
ref. \cite{sandu}.
We have benefitted from useful discussions with C. Capan. P. Gegenwart and P. Thalmeier.
B. D. acknowledges the hospitality and support of the Max Planck Institute for Chemical Physics of Solids, Dresden,
where part of this work was done.
This work was 
supported by the Magyary Zolt\'an postdoctoral
program of Foundation for Hungarian Higher Education and Research (AMFK) and by
the Hungarian
Scientific Research Fund under grant numbers OTKA TS040878, TS049881 T046269.

\bibliographystyle{apsrev}
\bibliography{ceco}

\end{document}